\documentclass[twocolumn,tighten]{aastex63}

\usepackage{amsmath}
\usepackage{xcolor}
\usepackage{wasysym}

\definecolor{my_color}{HTML}{3a18b1}

\definecolor{new_color}{HTML}{CF0000}
\definecolor{new_black}{HTML}{000000}
\usepackage{xfrac}

\newcommand{\Kepler}{{\it Kepler}}

\newcommand{\thisstar}{HR~8799}
\newcommand{\thisplanetb}{HR~8799~b}
\newcommand{\thisplanetc}{HR~8799~c}
\newcommand{\thisplanetd}{HR~8799~d}
\newcommand{\thisplanete}{HR~8799~e}

\newcommand{\msini}{\ensuremath{m\sin{i}}}

\newcommand{\be}{\begin{equation}}
\newcommand{\ee}{\end{equation}}

\newcommand{\mj}{M$_{\rm J}$}

\newcommand{\kms}{\ensuremath{\rm km\,s^{-1}}}

\submitjournal{ApJL}

\shorttitle{HR 8799 Exomoon Limits}
\shortauthors{Vanderburg \& Rodriguez}

\begin{document}

\title{First Doppler Limits on Binary Planets and Exomoons in the HR 8799 System }

\author[0000-0001-7246-5438]{Andrew Vanderburg}
\affiliation{Department of Physics and Kavli Institute for Astrophysics and Space Research, Massachusetts Institute of Technology, 77 Massachusetts Avenue, Cambridge, MA 02139, USA}
\author[0000-0001-8812-0565]{Joseph E. Rodriguez} 
\affiliation{Department of Physics and Astronomy, Michigan State University, East Lansing, MI 48824, USA}

\correspondingauthor{Andrew Vanderburg}
\email{andrewv@mit.edu}



\begin{abstract}

We place the first constraints on binary planets and exomoons from Doppler monitoring of directly imaged exoplanets. We model radial velocity observations of HR~8799~b,~c,~and~d from \citet{ruffio} and determine upper limits on the \msini\ of short-period binary planets and satellites. At 95\% confidence, we rule out companions orbiting the three planets more massive than $\msini = 2$ \mj\ with orbital periods shorter than 5 days. We achieve our tightest constraints on moons orbiting HR 8799 c, where with 95\% confidence we rule out out edge-on Jupiter-mass companions in periods shorter than 5 days and edge-on half-Jupiter-mass moons in periods shorter than 1 day. These radial velocity observations come from spectra with resolution 20 times lower than typical radial velocity instruments and were taken using a spectrograph that was designed before the first directly imaged exoplanet was discovered. Similar datasets from new and upcoming instruments will probe significantly lower exomoon masses. 

\end{abstract}



\section{Introduction}\label{introduction}

Three decades of exoplanet discovery have resulted in the detection of over 4500 planets outside the solar system\footnote{\url{https://exoplanetarchive.ipac.caltech.edu/cgi-bin/TblView/nph-tblView?app=ExoTbls&config=PS&constraint=default_flag=1}, accessed 18 September, 2021.} \citep{akeson} and the knowledge that planets outnumber stars in our galaxy \citep[e.g.][]{swift2013, dressingcharbonneau}. However, despite the prevalence of planets in our galaxy and the ubiquity of moons orbiting the planets in our solar system, astronomers have not yet securely detected any exomoons and have only a handful of unconfirmed candidates (e.g. \citealt{teacheykipping, lazzoni, limbach}, but see also \citealt{kreidbergnomoon}). Given the importance of Earth's moon on our planet's spin dynamics \citep{gongjie} and the potential habitability of icy moons in the outer solar system \citep[e.g.][]{reynolds}, continuing to search for moons beyond our solar system is a worthwhile endeavor.

There are two likely reasons for the lack of secure exomoon detections. First, large exomoons and binary planets (defined here as objects orbiting planets with enough mass that the system's center of mass is outside the primary planet, \citealt{stern2002}) are apparently rare. The \Kepler\ mission would have be highly sensitive to any nearly equal-sized binary-planet companions, but detected none \citep{lewis, teachey2018}. In a number of cases, upper limits on the presence of exomoons and binary-planet companions have been placed using transit photometry that rule out moons larger than Earth, and in the best cases rule out moons almost as small as Ganymede \citep{kippingmdwarfs, kipping40systems}. The rarity of large exomoons can be understood by the fact that moons forming circumplanetary disks are expected to have masses smaller than $10^{-4}$ times that the planetary host \citep{canupward}, while the gravitational capture of giant moons or binary planets is likely to be a rare dynamical event \citep{Ochiai}. Second, our observations are not yet sensitive to small moons. There are over 200 moons orbiting the 8 major Solar System planets and many more orbiting smaller bodies, but none of these moons would be detectable from afar with our current technology. Many different exomoon detection methods have been proposed \citep[e.g.][]{sartoretti1999, hook, cabreraschneider, lewispulsar, kippingmethod, agol2015, hellertransits, senguptamarley, forgan, hwang, vanderburg2018, lazzoni} but so far only transit photometry \citep{kipping40systems, teachey2018} and direct imaging \citep{lazzoni} have yielded constraints on the presence of exomoons. Future space observations from the Roman Space Telescope microlensing survey \citep{Spergel:2013, Penny:2019} are expected to either detect or constrain planets/moons orbiting low-mass brown dwarfs or free-floating planets, although unambiguous detections will be challenging \citep{hwang}. 

In this work, we place the first limits on exomoons and binary planets from radial velocity measurements of directly imaged exoplanets. This method has been mentioned occasionally in literature since 2005 \citep{hook, cabreraschneider, hellertransits, lillobox} and was recently discussed in detail by \citet{vanderburg2018} who showed that existing and forthcoming instruments could yield strong constraints on moons orbiting bright directly imaged planets. We apply this method to three such planets in the \thisstar\ multi-planet system \citep{marois1, marois2}. \thisstar\ hosts four directly imaged super-Jovian exoplanets named \thisplanete,{ \thisplanetd, \thisplanetc}, and \thisplanetb\ (in order of increasing orbital semimajor axis). \thisstar\ c, d, and e all have masses of $7.2_{-0.7}^{+0.6}$\mj, while the outermost planet \thisplanetb\ has a slightly lower mass of $5.8 \pm 0.5$\mj\ \citep{wang2018mass}. Recently, \citet{ruffio} observed \thisstar\ b, c, and d and reported repeated planetary radial velocity measurements with a typical precision of 1 \kms. Here, we show that these radial velocity observations are sensitive to massive exomoons in short-period orbits and use the time series to place upper limits on the presence of exomoon and binary companions to the \thisstar\ planets. This \textit{Letter} is organized as follows: Section \ref{observationsanalysis} describes the observations and our radial velocity analysis, Section \ref{results} describes the constraints we are able to place, and Section \ref{discussion} discusses the implications of our upper limits and the prospects for improved constraints or exomoon detections with future datasets.

\section{Observations and Analysis}\label{observationsanalysis}
\subsection{Keck OSIRIS Radial Velocities}\label{observations}
\begin{figure*}[htb] 
   \centering
   \includegraphics[width=\linewidth]{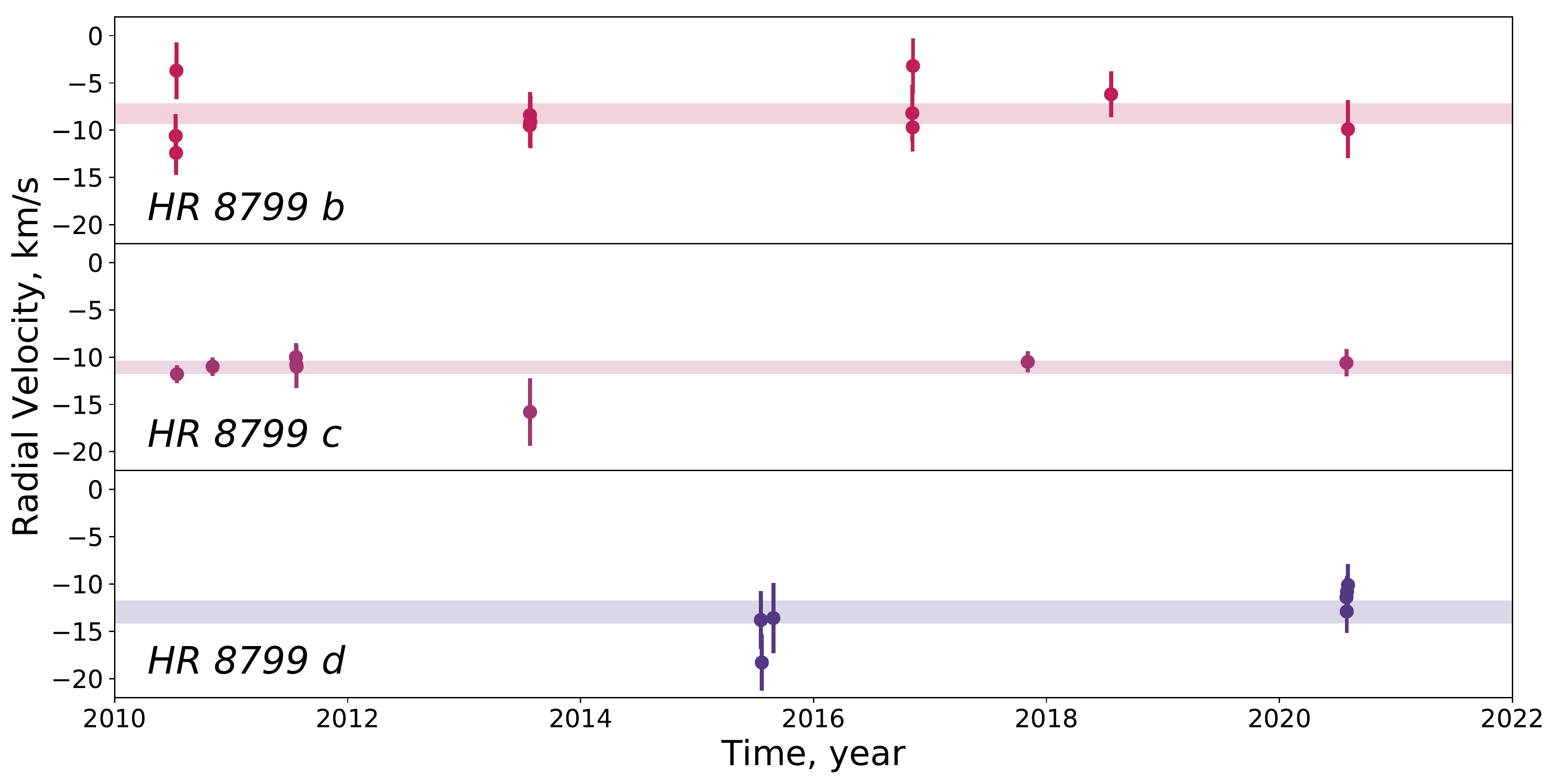} 
   \caption{Radial velocity observations of the \thisstar\ planets: \thisplanetb\ (\textit{top}), \thisplanetc\ (\textit{middle}), and \thisplanetd\ (\textit{bottom}). The points show the radial velocity measurements from \citet{ruffio} with their uncertainties inflated by adding the jitter term from our analysis in Section \ref{analysis} in quadrature. The transparent rectangular regions show the best-fit mean radial velocities from our analysis for each planet and their 1$\sigma$ uncertainties.}
   \label{rvs}
\end{figure*}

We use the radial velocity observations of \thisstar\ b, c, and d reported by \citet{ruffio}. These observations were made over the course of a decade (between UT 2010 July 11 and UT 2020 August 03) with the OH-Suppressing Infrared Integral Field Spectrograph (OSIRIS, \citealt{osiris, osirisgrating}) on the Keck telescopes. OSIRIS is a moderate-resolution ($\lambda/\Delta\lambda \approx 4000$) integral field spectrograph capable of measuring the spectra of multiple objects across a small field of view. While almost all observations were taken with the smallest plate scale (with 0\farcs02 pixels and a 0\farcs32$\times$1\farcs28 field of view),  one was taken with a slightly larger plate scale (0\farcs035 pixels and a 0\farcs56$\times$2\farcs24 field of view). Most of the observations reported were made in the K-band, some were derived from both H- and K-bands, and others were made entirely from the H-band observations (see Table \ref{tab:RVs}). \citet{ruffio} extracted spectra and measured radial velocities using least-squares template fitting, while simultaneously fitting for stellar contamination and modeling telluric absorption as a sum of vectors from a principal component analysis of many spectroscopic observations. The radial velocity calibration was performed by measuring the location of OH$^-$ sky emission lines.  In total, \citet{ruffio} reported 11 observations of \thisplanetb, 8 observations of \thisplanetc, and 7 observations of \thisplanetd. We list the these radial velocity measurements in Table \ref{tab:RVs} and show them in Figure \ref{rvs}. 

Two other groups have reported radial velocity measurements of the \thisstar\ planets using different instruments on the Keck telescopes. \citet{jiwang2018} used the NIRSPEC near infrared spectrograph behind the Keck adaptive optics system to measure the radial velocity of {\thisplanetc\ }and \citet{wang2021} used NIRSPEC with the the Keck Planet Imager and characterizer (KPIC) front-end to measure the radial velocity of all four planets in the HR 8799 system. Both groups achieved radial velocity uncertainties as small as 1-2 km/s, at the level of the OSIRIS observations used here, but only reported one measurement per planet. Because the different instruments used to collect the observations and the different data analysis methods likely introduce offsets between these measurements and the OSIRIS datasets, we do not include them in our analysis. 

\begin{table}[!h]
\centering
\caption{Radial Velocities of \thisstar\ b, c, and d}
\scriptsize
\begin{tabular}{lccccc}
\hline
Planet & Time & RV  & $\sigma_{\rm RV}$& Pixel&Filter/\\
Name & BJD & $\rm{km}\,\rm{s}^{-1}$ &$\rm{km}\,\rm{s}^{-1}$& Scale(\arcsec)&Band\\

\hline
\thisplanetb & 2455389.05 & -10.6 & 1.1 & 0.02 & K\\
\thisplanetb & 2455390.04 & -12.4 & 1.1 & 0.02 & K\\
\thisplanetb & 2455391.12 & -3.7 & 2.2 & 0.02 & H\\
\thisplanetb & 2456499.05 & -9.5 & 1.0 & 0.02 & K\\
\thisplanetb & 2456500.0 & -8.4 & 1.3 & 0.02 & K\\
\thisplanetb & 2456501.01 & -9.1 & 1.9 & 0.02 & K\\
\thisplanetb & 2457698.82 & -8.2 & 2.2 & 0.02 & K\\
\thisplanetb & 2457699.82 & -9.7 & 1.5 & 0.02 & K\\
\thisplanetb & 2457700.84 & -3.2 & 2.1 & 0.02 & K\\
\thisplanetb & 2458322.02 & -6.2 & 1.3 & 0.035 & K\\
\thisplanetb & 2459064.96 & -9.9 & 2.3 & 0.02 & K\\

\hline
\thisplanetc & 2455393.05 & -11.8 & 0.7 & 0.02 & K\\
\thisplanetc & 2455504.81 & -11.0 & 0.7 & 0.02 & H/K\\
\thisplanetc & 2455766.04 & -10.0 & 1.3 & 0.02 & K\\
\thisplanetc & 2455767.06 & -10.8 & 2.1 & 0.02 & H/K\\
\thisplanetc & 2455768.07 & -11.0 & 2.2 & 0.02 & H/K\\
\thisplanetc & 2456500.08 & -15.8 & 3.5 & 0.02 & K\\
\thisplanetc & 2458060.75 & -10.5 & 0.9 & 0.02 & H/K\\
\thisplanetc & 2459060.0 & -10.6 & 1.3 & 0.02 & K\\
\hline
\thisplanetd & 2457224.01 & -13.8 & 2.8 & 0.02 & K\\
\thisplanetd & 2457227.02 & -18.3 & 2.7 & 0.02 & K\\
\thisplanetd & 2457263.04 & -13.6 & 3.5 & 0.02 & K\\
\thisplanetd & 2459060.09 & -11.4 & 1.6 & 0.02 & K\\
\thisplanetd & 2459061.06 & -12.9 & 1.9 & 0.02 & K\\
\thisplanetd & 2459062.04 & -10.8 & 1.1 & 0.02 & K\\
\thisplanetd & 2459065.05 & -10.1 & 1.8 & 0.02 & K\\
\hline
\end{tabular}
\label{tab:RVs}
\begin{flushleft}
 \footnotesize{
  These observations are as reported by \citet{ruffio}.
 }
\end{flushleft}
\end{table}

\subsection{MCMC Radial Velocity Analysis}\label{analysis}

{We searched the radial velocity time series of all three planets for evidence of orbiting companions. We calculated Lomb-Scargle periodograms \citep{lomb, scargle} of the three time series, and saw no clear peaks at periods between 0.1 and 10 days. A boot-strap analysis yielded false alarm probabilities for the highest peaks in the power spectra for \thisplanetb, c, and d of 68\%, 49\%, and 15\%, respectively. We concluded that there is no evidence for orbiting companions in the OSIRIS radial velocity observations. }

We therefore determined upper limits on the presence of putative companions to the \thisstar\ planets using Markov Chain Monte Carlo (MCMC) explorations of plausible short-period orbits. Radial velocity observations are unable to constrain the true mass of planetary companions because of a well-known degeneracy with the companion's orbital inclination, $i$. We therefore place constraints on \msini\ rather than the true masses of orbiting binary planets or moon. In our MCMC fits, we modeled each planets' radial velocity time series as the sum of a single Keplerian signal describing a moon/binary-planet companion (restricted to circular orbits), a constant radial velocity offset, and a linear velocity trend in time (to model each planets' orbital motion). We fit for a white noise ``jitter'' term, $j$, which we added in quadrature to the error bars, $\sigma$, reported by \citet{ruffio} to account for any additional sources of radial velocity uncertainty. That is, the effective uncertainty, $\sigma_{\rm new},$ of each observation was inflated to $\sigma_{\rm new}\equiv \sqrt{\sigma^2+j^2}$. We imposed a Gaussian prior on the masses of the three planets with central values and widths for \thisstar\ b, c, and d of $M_b = 5.8 \pm 0.5$,  $M_c = 7.2 \pm 0.7$, and $M_d = 7.2 \pm 0.7$ \citep{wang2018mass}. The radial velocity datasets from \citet{ruffio} are sparsely sampled, so we were forced to restrict the putative companion orbital periods to be between 0 and 10 days. This restricted parameter space is still interesting though; more than a quarter of the moons in the solar system, comprising nearly half the total moon mass, orbit in periods shorter than 10 days\footnote{\url{https://ssd.jpl.nasa.gov/sats/elem/}}. {We forced the \msini\ of the putative companion to be nonnegative with a uniform prior over the interval $[0,\infty)$.}  Finally, we restricted the orbital phase of conjunction to be between 0 and 1. 

All in all, our model included seven parameters: the mass of each host planet, the \msini, orbital period, and phase of conjunction for the putative moon/binary-planet companion, a radial velocity offset, a linear radial velocity trend, and an excess radial velocity ``jitter'' term.   We explored this seven-dimensional parameter space using \texttt{edmcmc} {\citep{edmcmc}}, a Python implementation\footnote{\url{https://github.com/avanderburg/edmcmc}} of the differential evolution MCMC sampler of \citet{terbraak}. For each planet, we ran 50 parallel chains for 400,000 links, discarding the first 20,000 links as burn-in. We assessed convergence using the Gelman-Rubin diagnostic \citep{gelmanrubin}, which was below 1.02 for all parameters.  As an independent check on our analysis, we ran a similar MCMC fit for \thisplanetb\ using the EXOFASTv2 software \citep{eastman2019} and found consistent results. 

\begin{figure*}[htb] 
   \centering
   \includegraphics[width=\linewidth]{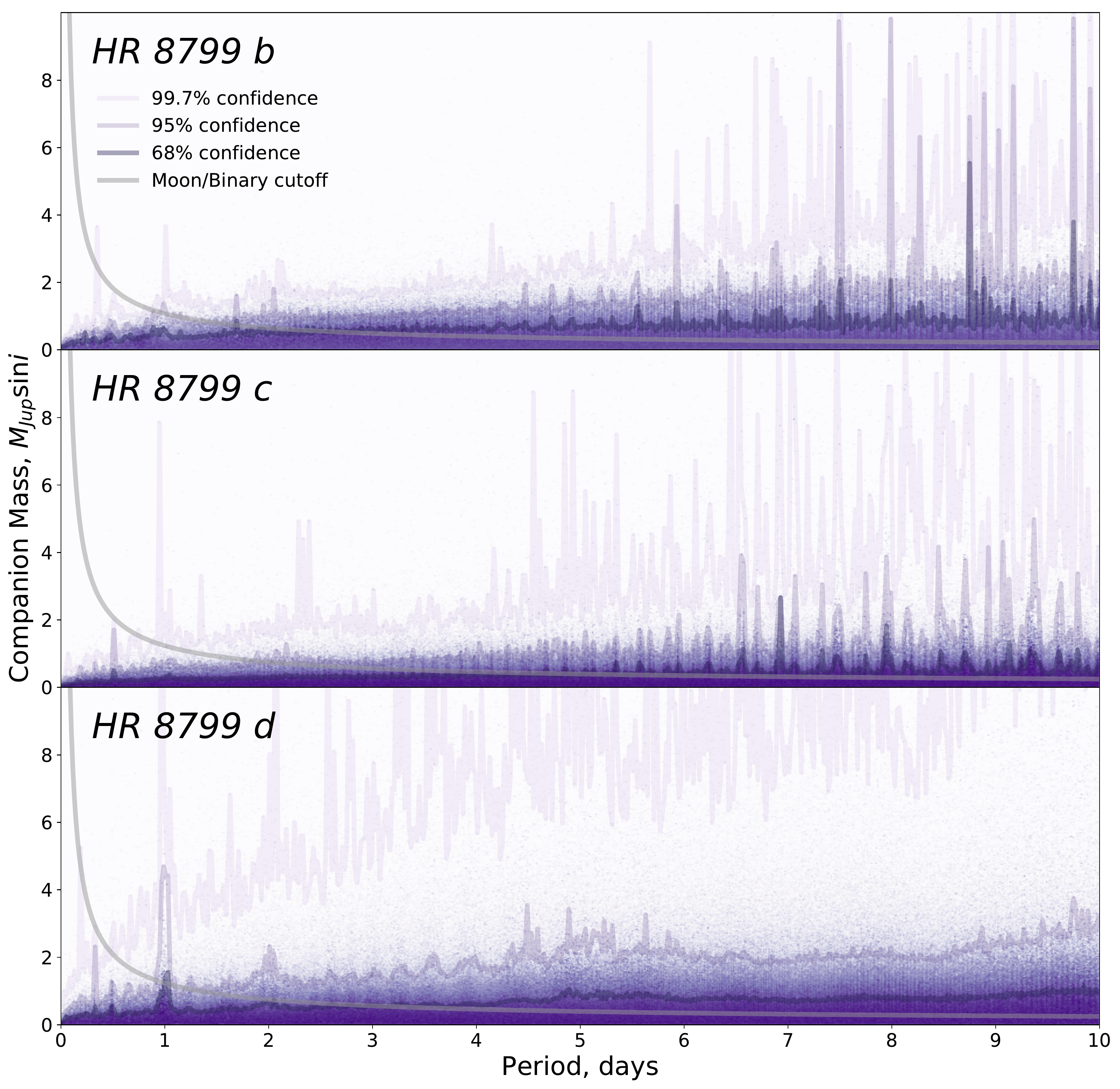} 
   \caption{Limits on the presence of orbiting companions to the \thisstar\ planets: \thisplanetb\ (\textit{top}), \thisplanetc\ (\textit{middle}), and \thisplanetd\ (\textit{bottom}). The background shading shows the posterior probability distribution of allowed moon/binary planet \msini\ as a function of orbital period. The dark, medium, and light purple lines show the $1\sigma$, $2\sigma$, and $3\sigma$ (68\%, 95\%, and 99.7\% confidence) contours as a function of orbital period. The grey lines show the boundary between a binary-planet system and a planet/moon system based on the definition of \cite{stern2002} -- objects below the grey lines are moons, while objects above the grey lines are binary-planet companions. The spiky vertical structures in the probability distribution and contours at longer orbital periods are due to the dataset's sparse sampling.}
   \label{constraints}
\end{figure*}

\section{Results}\label{results}

Once our MCMC analysis was complete, we marginalized over most of our free parameters to yield constraints on the presence of exomoons and binary-planet companions as a function of moon \msini\ and orbital period. Our constraints are summarized in Figure \ref{constraints}, and our $1\sigma$, $2\sigma$, and $3\sigma$ limits are provided as Data Behind the Figure. 

We find that for all three planets, we are able to rule out most companions with \msini\ greater than 2\mj\ in periods shorter than 5 days. In particular, our 95\% confidence limits for companions in periods shorter than 5 days are 1.1 \mj, 0.82 \mj, and 1.7 \mj\ for \thisstar\ b, c, and d respectively. Between 5 and 10 days, our results are less constraining: we rule out companions more massive than \msini\ = 2.4 \mj, 1.8 \mj, and 2.3 \mj\ for the three planets at 95\% confidence. Our constraints are stronger for short-period companions for two reasons: 1) the radial velocity signal scales with period as $P^{-1/3}$ so longer-period companions induce smaller signals and 2) the sparse sampling of the RV observations causes poor phase coverage at longer orbital periods, making it more difficult to rule out all companions. The effects of the sparse radial velocity sampling is visible in Figure \ref{constraints} as vertical spikes in both the posterior probability distribution and our $1\sigma$, $2\sigma$, and $3\sigma$ limits at specific orbital periods. The spikes become more prominent at periods longer than 5 days, reflecting the typical sampling of the dataset, which have observations on a few consecutive nights interspersed between years-long gaps without data.

To determine whether our limits predominantly rule out exomoons versus binary planets, we calculated the boundary between these two classes of objects in \msini/period space (assuming edge-on inclinations). \citet{stern2002} define a moon as an orbiting companion with a small enough mass and semimajor axis that the moon/planet system center of mass remains within the radius of the primary planet. A binary planet is any companion massive enough that the system center of mass lies outside the primary planet. Using this rule, we calculate the mass boundary, $M_{\rm boundary}$ above which a companion is a binary planet rather than a moon: 

\begin{equation}
    M_{\rm boundary} = \frac{M_p\,R_p}{a - R_p}
\end{equation}

\noindent where $M_p$, $R_p$, and $a$ are the planet mass, planet radius, and companion semimajor axis, respectively. We plot this curve in \msini/period space in Figure \ref{constraints} and find that our radial velocity constraints are sufficient to rule out almost all edge-on binary planets and some massive edge-on exomoons with periods shorter than 1-2 days. At orbital periods longer than about 2 days, our observations are only sensitive to binary planets. 

A natural by-product of our MCMC analysis is that we measure the mean velocity of each planet over the span of the observations. Since we also fit for a linear velocity trend, we calculate the velocity of each planet at the mean time of all of its observation timestamps. We measure mean velocities of $-8.4 \pm 1.7$~\kms, $-11.1 \pm 1.1$~\kms, and $-13.0 \pm 1.8$~\kms\ for \thisstar\ b, c, and d, respectively. These velocities are largely consistent with the values ($-9.1 \pm 0.4$~\kms, $-11.1 \pm 0.4$~\kms, and $-11.6 \pm 0.8$~\kms\ for \thisstar\ b, c, and d) calculated by \citet{ruffio}, but have larger error bars (due to our use of a jitter term to inflate the individual error bars) and are slightly more extreme (likely because including the jitter term gives slightly more relative credence to apparently outlying datapoints). The more extreme velocities we measure for the planets match the predicted velocities from orbital fits ($-8.6^{+0.5}_{-0.6}$~\kms, $-10.5^{+0.5}_{-0.6}$~\kms, and $-13.2^{+0.6}_{-0.7}$~\kms\ for \thisstar\ b, c, and d, \citealt{wang2021}) slightly better than the \citet{ruffio} averages, indicating that this type of jitter term may be a useful model of the dataset's noise properties. 

Another by-product of our MCMC analysis is an estimate of the radial velocity ``jitter,'' or excess radial velocity scatter in each dataset. We measure an excess radial velocity uncertainty of $2.1^{+1.3}_{-1.0}$~\kms, $0.7^{+1.0}_{-0.5}$~\kms, and $1.3^{+1.9}_{-0.9}$~\kms\ for the \thisstar\ b, c, and d observations respectively. For both \thisstar\ c and d the most probable value of the jitter value is 0, but the \thisplanetb\ dataset shows a preference for non-zero jitter. This likely because \thisstar\ b is the faintest of the \thisstar\ planets, and therefore is the most susceptible to low-level residual systematic errors in the spectral extraction or the telluric and starlight correction. 

\section{Discussion}
\label{discussion}

It is not particularly surprising that the \thisstar\ planets do not have detectable short-period Jovian moons or binary-planet companions. Among the thousands of known exoplanets, there are still no securely detected exomoons or binary planets, even though companions the mass and size of Jupiter should have been detectable around most \Kepler\ target stars \citep{lewis}. Nevertheless, our constraints ruling out massive edge-on exomoons and binary planets around \thisstar\ are novel. Almost all published limits on the presence of exomoons \citep[e.g.][]{kippingmdwarfs, kipping40systems} come from transiting exoplanet systems, which due to selection effects orbit close to their host stars (within 1-2 AU). At these short orbital periods, the planets' Hill spheres are smaller, limiting the parameter space in which moons can survive. Indeed, only three of the hundreds of known moons in the solar system are found within 2 AU of the Sun. So far, only \citet{lazzoni} have published constraints on moons orbiting planets at similar orbital radii to the \thisstar\ planets, and their direct imaging constraints only probe moons orbiting far from their host planets. Our constraints are first probes of moons and binary planets in a new parameter space {(planet semimajor axis $\gtrsim$ 10 AU and moon orbital period $\lesssim$ 10 days).} 

One limitation of the radial velocity method of exomoon (and exoplanet) detection is that it is most sensitive to companions with close to edge-on orbital inclinations.  The radial velocity signal induced by a companion is proportional to the sine of its orbital inclination, and generally it is not possible to measure a companion's true mass without an independent inclination measurement. It is therefore possible that massive exomoons or binary planets are hiding around the \thisstar\ planets in nearly face-on orbits. The \thisstar\ planets orbit their host star with orbital inclinations of about 20 degrees; if moons or binary planets in the system orbit in the same plane, their radial velocity signals would be attenuated by about a factor of 3. However, such massive moons and binary planets likely must have formed by dynamical capture, and would not necessarily orbit in the same plane as the planetary orbits. 

Another promising way to detect exomoons around directly imaged planets is by measuring the astrometric wobble of the planet due to an orbiting satellite \citep{agol2015}. Recently, the GRAVITY instrument on the Very Large Telescope Interferometer (VLTI) has achieved extremely precise astrometric measurements (uncertainties ranging from 20-150 microarcseconds) of directly imaged planets \citep{gravitybetapic1, gravitybetapic2, gravitybetapic3}, including \thisplanete\ \citep{gravityhr8799e}. The astrometric signal induced by a Jupiter-mass companion in a 10 day orbit around one of the \thisstar\ planets would be about 60 microarcseconds, roughly 3 times larger than the smallest single-point GRAVITY astrometric uncertainties. Since the radial velocity semiamplitude of a Jupiter-mass companion in a 10 day orbit around one of the \thisstar\ planets ($\approx 2.4~\kms$) is also about 3 times larger than the best smallest single-point OSIRIS radial velocity uncertainty measurement, astrometry and radial velocities are likely to be similarly sensitive to companions at 10 day orbital periods. These techniques are highly complementary because a companion's radial velocity amplitude decreases with orbital period (radial velocity semiamplitude $k \propto P^{-1/3}$), while the astromeric amplitude increases with orbital period (astrometric amplitude $\theta \propto P^{2/3}$). So while radial velocities will likely provide the tightest constraints on close-in exomoons, astrometry could be a more sensitive probe of long-period companions. {Owing to its extremely high angular resolution, GRAVITY can spatially resolve the Hill spheres of many nearby directly imaged exoplanets \citep{wang2021pds}, including those around \thisstar. It therefore may also be possible to use GRAVITY to directly detect luminous orbiting exomoons and binary-planet companions on wide orbits around their primary planet.} Detecting exomoons orbiting directly imaged planets using the transit method \citep[e.g.][]{cabreraschneider} will be very challenging given the need for high-duty cycle observations on 10 meter class telescopes (required to resolve the individual planets). However, targeted observations following up on known signals from either radial velocities or astrometry could be feasible.  

Perhaps the most interesting result of this work is not the limits on exomoons and binary planets themselves, but the fact that it was possible to constrain the presence of companions in the first place. Detecting exoplanets around stars with the radial velocity method requires many high signal-to-noise observations with stabilized high-resolution spectrographs, but the observational requirements for detecting exomoons are dramatically lessened due to the low mass of the host planet. This is why it was possible for us to place meaningful constraints on the presence of exomoons using a moderate resolution spectrograph ($\lambda/\Delta\lambda \approx 4000$) that was designed \citep{larkin2003} before the discovery of the first directly imaged exoplanet \citep{chauvin}. A new generation of specially designed instruments for taking high resolution spectra of directly imaged planets will dramatically improve the detection limits. Doppler uncertainties scale with the full width at half maximum of spectral lines to the 1.5 power \citep{lovisfischer}, so increasing the spectral resolution from $\lambda/\Delta\lambda \approx 4000$ (with OSIRIS) to $\lambda/\Delta\lambda \approx 40000$ (with an instrument like KPIC) with signal-to-noise held constant could decrease radial velocity uncertainties by a factor of 30. Upgrades to existing instruments like {KPIC/Keck \citep{jovanovic}, CRIRES+/SPHERE/HiRISE/VLT \citep{dorn2014, vigan}, and IRD/SCExAO/REACH/Subaru \citep{kotani}} and new instruments on 30 meter class telescopes like GMTNIRS/GMT \citep{jaffe} and HIRES/ELT \citep{elthires} will yield high signal-to-noise spectra, potentially enabling Doppler detections of rocky exomoons.

\acknowledgments

We thank Jason Eastman for helpful discussions.  {We thank the anonymous referee for a quick and constructive report and for the idea that GRAVITY's high angular resolution could allow it to actually resolve luminous moons and binary-planet companions}. This research has made use of NASA's Astrophysics Data System, the NASA Exoplanet Archive, which is operated by the California Institute of Technology, under contract with the National Aeronautics and Space Administration under the Exoplanet Exploration Program, and the SIMBAD database, operated at CDS, Strasbourg, France.

%

\facilities{Keck I (OSIRIS)}


\software{matplotlib \citep{plt},
          numpy \citep{np}, scipy \citep{scipy}, EXOFASTv2 \citep{eastman2019}, {edmcmc \citep{edmcmc}}
          }



\vspace{5mm}





\end{document}